\tolerance 500
\magnification=\magstep1
\rightline{TAUP 2378-96}
\bigskip
\centerline{\bf Lax-Phillips Theory and Quantum Evolution}
\vskip 0.75truecm
\centerline{  L.P.
 Horwitz$^{\S\dag}$, E. Eisenberg$^{\S}$ and Y. Strauss$^{\dag}$}
\smallskip
\centerline{$^{\S}$  Department of Physics, Bar-Ilan University}
\centerline{ Ramat-Gan 52900, Israel}
\centerline{$^{\dag}$  School of Physics and Astronomy,}
\centerline{Raymond and Beverly Sackler Faculty of Exact Sciences }
\centerline { Tel-Aviv University, Ramat-Aviv, Israel}
\bigskip
\noindent
{\it Abstract:}\   The scattering theory of Lax and
Phillips, designed primarily
for hyperbolic systems, such as electromagnetic or acoustic
waves, is described.  This theory provides a realization of the
theorem of Foias and Nagy; there is a subspace of the Hilbert
space in which the unitary evolution of the system, restricted
to this subspace, is realized as a semigroup.  The embedding of
the quantum theory into this structure, carried out by Flesia and
Piron, is reviewed.  We show how the density matrix for an effectively
pure state can evolve to an effectively mixed state (decoherence) in
this framework.  Necessary conditions are given for the realization
of the relation between the spectrum of the generator of the semigroup
and the singularities of the $S$-matrix (in energy representation).
It is shown that these conditions may be met in the Liouville space
formulation of quantum evolution, and in the Hilbert space of
relativistic quantum theory.
\bigskip
\noindent{\bf 1. INTRODUCTION}
\smallskip
\par The unstable quantum system is an important example of
irreversible phenomena in nature. Such systems, ranging
from excited atomic states to short-lived elementary
particles, are characterized by what is generally observed
to be an irreversible evolution. These phenomena raise the
question of explanation of such processes from first
principles. Moreover, since most of the decay processes
are observed experimentally to obey an exponential decay
law, one expects this behavior to follow from very general
assumptions.
\par Irreversible evolution in the quantum
theory has been described by the addition of non-Hermitian
terms to the Hamiltonian, such that it has complex
eigenvalues, and the induced evolution is non-unitary.
Structures of this type were originally introduced by
Gamow [1] who studied the effect of assigning complex
eigenvalues to the energy spectrum, and hence introduced a
kind of generalized eigenvector. Wu and Yang [2]
 parameterized the $K$-meson decay in this way. In
this method, the non-Hermitian terms in the Hamiltonian
are introduced phenomenologically, and may only indirectly
be associated with some known interaction in a more fundamental
Hamiltonian.
\par Weisskopf and Wigner [3], in a well known paper in 1930,
 introduced an alternative approach to the decay
problem. According to their approach, the evolution takes
place in a Hilbert space which is a direct sum of two
subspaces: the subspace of the decaying states and that of
decay products. These two subspaces are stable under the
``free'' evolution induced by $H_0$, but are combined
linearly under the full evolution induced by
$H\,=\,H_0+V$. In this Hilbert space, the evolution is
unitary, and hence its generator, i.e., the Hamiltonian,
is self-adjoint. The decay is described as the probability
flow from the subspace of the decaying states to its
complement, the subspace of the decay products.
They studied perturbatively, for the single-channel case, what
has become known as the {\it survival amplitude}
$$A(t)\,=\,(\psi,e^{-iHt}\psi)\,,\eqno(1.1)$$ which is the
probability amplitude for the system to remain
 in the discrete state until
time $t$.
In the following we will describe
this approach, and pose critical problems, motivating the
development of a more general theory.
\par  Let us denote the
projection operators on these two subspaces as $P$ and ${\bar
P}$, such that $P+{\bar P}=1$. For the decay problem, the
basic quantity is the {\it reduced motion}
$$U'(t)\,=\,PU(t)P\,.\eqno(1.2)$$ where
$U(t)\,=\,e^{-iHt}$, which governs the time evolution of
the subspace $P{\cal H}$ of the unstable states. From this
one can derive the decay law of the unstable states. If
$\{\phi_i\}$ is an orthonormal basis of $P{\cal H}$, the
probability that an unstable state $\phi$, which exists at
time $t=0$, is in the subspace $P{\cal H}$ of unstable
states at time $t$ is given by $$p(t)\,=\,\sum_i
|(\phi_i,U(t)\phi)|^2\ =  {\rm Tr} (U'(t)^\dagger U'(t)P_\phi),
\eqno(1.3)$$
where $P_\phi = \vert \phi \rangle \langle \phi \vert $.
\par The total evolution operator $U(t)\,=\,e^{-iHt}$, and the
resolvent $R(z)\,=\,(z-H)^{-1}$ are related to each other
by the (inverse) Laplace transform $$U(t)\,=\,{1\over 2\pi
i}\oint R(z)e^{-izt}dz\,,\eqno(1.4)$$ where the
integration contour is around the spectrum of $H$. If we
project this operator into the subspace $P{\cal H}$, we
can obtain a similar relation which expresses the reduced
motion $U'(t)$ in terms of the {\it reduced resolvent}
$R'(z)\,=\,PR(z)P$: $$U'(t)\,=\,{1\over 2\pi i}\oint
R'(z)e^{-izt}dz\,.\eqno(1.5)$$
\par By differentiating Eq. $(1.3)$ and setting $t=0$, one
sees that the initial decay rate is necessarily zero (providing
that the Hamiltonian is defined on the initial state); in fact,
it is easy
to show that the change in $p(t)$ is O$(t^2)$.  The intermediate
and long time behavior follow most simply by an examination of
the relation $(1.5)$.  Deforming the contour of integration which
runs below the real positive spectrum of $H$ to the negative
imaginary axis, where its contribution for large times is small,
the remaining contribution of the contour running above the
spectrum of $H$ can be estimated by bringing this contour
continuously through the cut.  When resonances exist, this
contour will pass through simple poles on the way to the negative
imaginary axis in the second sheet (as can be explicitly demonstrated
[4] in the Lee-Friedrichs model [5]).  The residues of the poles
may dominate the time dependence for intermediate times, and
give the approximate (due to the presence of residual contributions
from the integrals along the negative imaginary axis) exponential
decay behavior.  For very long times, the pole contributions
disappear, and the remaining integration around the branch cut
results in an inverse power law asymptotic behavior [6].
\par It is not difficult to see that an irreversible evolution
must be described by a semigroup [7] (for the reversible
case this is a group induced by a unitary transformation),
where we define a semigroup as follows:
\par Let $\{Z(t)\}$ be a family (over $t$) of operators on a Hilbert
space; then $Z(t)$ is an element of a semigroup if
$$ Z(t_1) Z(t_2) = Z(t_1 + t_2), \qquad t_1,t_2 \geq 0 \eqno(1.6)$$
The semigroup
  is said to be {\it strongly contractive} if $\Vert Z(t)\Vert
\rightarrow 0$, for $t \rightarrow \infty$,
 where $\Vert A\Vert$ is the operator norm of $A$.
On the other hand, it can be shown that the
reduced motion, as described above, cannot generate a
semigroup [8].
\par There is, furthermore, another, perhaps more fundamental
problem associated with the general method of Wigner and
Weisskopf; this is that the expression (1.1) for the
survival amplitude implicitly assumes the existence of a
linear superposition (we restrict our discussion here to the
one-channel case)
$$e^{-iHt}\psi\,=\,A(t)\psi\,+\,\chi(t)\,,\eqno(1.7)$$
where $\chi(t)$ represents the decayed system and
$(\psi,\chi(t))=0$. In general this linear superposition
does not correspond to any physical situation in our
experience; a short-lived particle, for example, is seen
as either the particle before the decay, or the decay
products at a certain time, which can not be predicted.
This linear superposition does not correspond to the
object that we see experimentally in such a process.
\par In the framework of the theory of Weisskopf and Wigner,
techniques have been developed which are capable of displaying
the exact semigroup behavior of an unstable system [9].  As we
have remarked above, the Lee-Friedrichs model [5] provides
a simple but useful example of an unstable system for which
the evolution equations are completely soluble [4].
The eigenvalue equation $Hf(z)= z f(z)$, for $f \in {\cal H}$
has a formal solution which does not, however, satisfy the
equation.  The condition to satisfy the equation coincides with
the condition for a pole in the resolvent $R'(z)$, and can
only be satisfied in the second Riemann sheet. The scalar product
of the eigenvalue equation with a vector for which the unperturbed
energy representation is analytic in a domain containing the
second sheet pole can be continued to the second sheet, and
at the pole position, the
equation is identically satisfied.  Since the ``eigenvector'' obtained
in this way is in the space dual to a set of vectors with this
restrictive analyticity requirement, corresponding to a subspace
of $\cal H$, it is an element in the large space of
a Gel'fand triple (rigged Hilbert space). The definition
of this vector depends on the domain of analyticity chosen, and
its physical interpretation is not clear, except for the fact that
its (extended) unitary evolution is that of an exact exponential
decay.
\par The problems discussed above are essentially related to the
attempt to
describe an unstable system in a framework more suitable
to the description of reversible phenomena. In what
follows we will show another approach to irreversible
phenomena which attempts to solve these difficulties.
\bigskip
\noindent
{\bf 2. LAX-PHILLIPS THEORY AND THE EXACT SEMIGROUP}
\smallskip
\par The characterization of a system undergoing an
irreversible process cannot, in principle, be specified at
a given instant of time. In fact, the physical quantities
describing such processes
 involve time measurements (that is, measurements of
the time at which certain defined phenomena occur).
Therefore, the information about the decay which is to be
deduced from the state is associated with its distribution
in time, an essential property of the system, just
as the location or momentum of a quantal particle. The
time variable is, from this point of view, an internal
degree of freedom of the system, which provides a
framework for the description of interactions which can
influence the structure of the state [10].
The dynamical evolution of the system involves a change in
its internal structure, including its distribution in $t$
along with other observables characterizing the state.
This evolution, parameterized by the laboratory time
$\tau$ (which is not a dynamical variable), is defined on
a Hilbert space ${\bar{\cal H}}$ with $t$ in its measure space
with norm given by (e.g., with Lebesgue measure) $$\int
\Vert \psi_t^{\tau}\Vert^2dt\,=\,\Vert \psi^{\tau}\Vert^2\, ,
\eqno(2.1)$$
where the norm in the integral is taken as the norm in ${\cal H}_t$,
a member of a
family of auxiliary Hilbert spaces (all isomorphic),
defined for each $t$.

\par The theory of Lax and Phillips [11], designed for systems
of hyperbolic differential equations describing the
scattering of, e.g., electromagnetic or acoustic waves,
and the Floquet theory [12] for periodic time dependent
quantum mechanical problems are examples of such a
structure. Piron [7] has shown that methods of this type
are applicable to the general time dependent quantum
mechanical problem. Recently, Flesia and Piron [13] have
shown that scattering problems in quantum theory can be
put in the form of Lax-Phillips theory (Horwitz and
Piron [14] have discussed its applicability to the
problem of the unstable system) by forming a direct integral
of the quantum mechanical Hilbert spaces ${\cal H}_t$ over
$t$ in order to construct a larger space ${\bar{\cal H}}$
which includes $t$ in its measure space.
\par Lax-Phillips theory [11] assumes the existence of a
one-parameter unitary group of evolution on a Hilbert
space $\bar{\cal H}$, and incoming and outgoing subspaces
$\cal D_-$ and $\cal D_+$ such that $$U(\tau){\cal
D}_+\,\subset\,{\cal D}_+\,,\,{\rm for\, all}\,\tau>0$$
$$U(\tau){\cal D}_-\,\subset\,{\cal D}_-\,,\,{\rm for
\,all}\,\tau<0$$ $$\bigcap_\tau U(\tau){\cal
D}_\pm\},=\,\{0\}$$
$${\overline{\bigcup_\tau U(\tau){\cal
D}_\pm}} \,=\,\bar{\cal H} \eqno(2.2) $$ where $\tau$ is
the evolution parameter identified with the laboratory
time. It follows from a theorem of Sinai
[15] that  $\bar{\cal H}$ can be foliated in
such a way that it can be represented as a family of
(auxiliary) Hilbert spaces in the form
$L^2(-\infty,+\infty; {\cal H}_t)$, over  Lebesgue
measure in $t$ , and all the  ${\cal H}_t$ are isomorphic
(we therefore sometimes refer to these spaces simply as
${\cal H}$) and determined up to unitary equivalence.
The scalar product in $\bar{\cal H}$ is given
by  $$(f\,,\,g)\,=\,\int_{-\infty}^\infty
(f_t\,,\,g_t)_{{\cal H}_t} dt \,. \eqno(2.3)$$

Lax and Phillips show that there are unitary operators
$W^{-1}_+,W^{-1}_-$ which map the elements of $\bar{\cal
H}$ into representations, called the outgoing and incoming
translation representations, for which the evolution is
translation in $t$. The subspaces  $\cal D_+,D_-$
correspond to the sets of functions with, in these
representations, support in semi-infinite segments of the
positive and negative $t$-axis respectively. They define
the $S$ matrix abstractly as the map from the incoming
translation representation to the outgoing one, i.e.,
$S=W_+^{-1}W_-$. This map is defined up to unitary
transformations on the auxiliary spaces $\{{\cal H}_t\}$,
and refers to the equivalence classes for which the
incoming and outgoing representations have the property
that the evolution is represented by translation.
\par Lax and Phillips furthermore define the operator
$${\cal Z}(\tau)\,=\,P_+U(\tau)P_-\eqno(2.4)$$
on ${\bar{\cal H}}$, where $P_{\pm}$ is the projection on
the orthogonal complement of $\cal D_{\pm}$. This operator
vanishes on $\cal D_{\pm}$ and maps the subspace $${\cal
K}\,=\,{\bar {\cal H}} \ominus ({\cal D}_+\oplus {\cal
D}_-)\,,\eqno(2.5)$$  into itself. These mappings form a
semigroup [11], i.e., for $\tau_1,\tau_2\geq 0$, $${\cal
Z}(\tau_1){\cal Z}(\tau_2)\,=\,{\cal
Z}(\tau_1+\tau_2)\,,\eqno(2.6)$$  and this semigroup is
strongly contractive, i.e., for each $\phi\in{\cal K}$ and
any $\epsilon$, there exists a $\tau_{\phi}$ such that
$$||{\cal Z}(\tau)\phi||_{\bar{\cal
H}}\,<\,\epsilon\eqno(2.7)$$  for $\tau>\tau_{\phi}$. It
can be shown that ${\cal
Z}(\tau)$ is just the unitary evolution $U(\tau)$
projected into the subspace $\cal K$. Since the states
which lie in the subspaces $\cal D_{\pm}$, in the case of
scattering, describe the incoming and outgoing waves which
are not influenced by the interaction, the states which
lie in $\cal K$ describe the unstable states, i.e.,
resonances of the scattering. From this point of view, the
Lax-Phillips semigroup is analogous to the reduced motion
discussed in the previous section.
\par This theory constitutes a constructive realization of the
theorem of Foias and Nagy [16], which states that given a
semigroup on a Hilbert space ${\cal H}$, there is a bigger
Hilbert space ${\bar{\cal H}}$ which contains it, in which
the evolution is a one-parameter unitary group, and this
unitary group restricted to $\cal H$ is that semigroup.\footnote
{*}{The theorem states that this construction is minimal.
We conjecture but have not proved that the Lax-Phillips
construction is minimal.  We thank G. Emch for a discussion of
this point.}
\par Flesia and Piron [13] have shown that the quantum theory may be
embedded in a Lax-Phillips theory by considering the
family of Hilbert spaces of the usual quantum theory on the
parameter $t$ as the auxiliary spaces of Lax and Phillips; the
large Hilbert space ${\bar{\cal H}}$ is then the direct integral
of these quantum  mechanical spaces over all values of the time
$t$ with Lebesgue measure.
 The form of the theory adopted by
Flesia and Piron [13] distinguishes the elements of these
equivalence classes, and constructs an $S$-matrix which
maps the auxiliary space in the incoming translation
representation to the auxiliary space of the outgoing one.
In the model that they use to illustrate this structure,
this map corresponds to a pre-asymptotic form of the
$S$-matrix of the usual scattering theory. Their model
assumes that the subspaces $\cal D_+,D_-$ are represented
in the ``free'' representation, for which the free
evolution is translation, by $L^2(-\infty,\rho_-;{\cal
H})$, $L^2(\rho_+, \infty;{\cal H})$, respectively. In the
limit in which the interval between the two semi-infinite
regions of support tends to infinity, their $S$-matrix
becomes the usual $S$-matrix. In this construction, Flesia and Piron
assume the form
$$\psi_{t+\tau}^{\tau}\,=\,W_t(\tau)\psi_t^0\,,\eqno(2.8)$$
where, since $W_t(\tau)$ represents an evolution, it follows that
$$W_{t+\tau_1}(\tau_2)W_t(\tau_1)\,=\,W_t(\tau_1+\tau_2)
\,.\eqno(2.9)$$
\par Lax and Phillips prove that the $S$-matrix
(in their construction) is a
multiplicative operator in the spectral representation of
the generator of the unitary evolution
$K$ (which is the Fourier transform of the translation
representation), i.e.,
$$(S\psi)_{\sigma}\,=\,S(\sigma)\psi_{\sigma},$$   and that the
eigenvalues of the generator of the semigroup ${\cal
Z}(\tau)$ correspond to the singularities of the analytic
continuation of $S(\sigma)$. The eigenstates corresponding
to these eigenvalues are analogous to the generalized
eigenstates found in the framework of Weisskopf and
Wigner, as discussed in Section 1. Thus, the
$S$-matrix contains all the information about the unstable
states . It can be seen [10], however,
that the $S$ matrix obtained from a model in which the evolution
is given in
the form (2.8) has no $t$-dependence, and hence its
spectral representation
is trivial.  In this form, one therefore has no relation between
the singularities of the $S$-matrix and the spectrum of the generator
of the semigroup.
\par Although the generalization of Lax-Phillips  theory by
Flesia and Piron [13]
provides a new point of view for scattering theory, we see
that to extend the theory further to include a description
of the evolution of an unstable system, it is necessary to
generalize the law of evolution to that of a nontrivial
integral operator over the time.
\par  The most general linear
evolution law has the form
$$(U(\tau)\psi)_{t+\tau}\,=\,\int_{-\infty}^{+\infty}
W_{t,t'}(\tau)\psi_{t'}dt' \, .  \eqno (2.10)$$
We shall show that this type of evolution, which goes
beyond the formulation of Flesia and Piron [13] and Floquet theory [12],
can correspond to unitary evolution in $\bar{\cal H}$ with
a nontrivial $S$-matrix for which the singularities of its
Fourier transform are associated with the spectrum of the
generator of the Lax-Phillips semigroup.
As we shall show below, the form of the evolution law (2.10) has
a natural realization in Liouville space as well as in the
framework of relativistic quantum theory.
\par Let us now study for this general evolution, some
properties of the $S$-matrix $$(S\psi)_t\,=\,\int
S_{t,t'}\psi_{t'}dt',$$ and show that in this general case
the S-matrix must have the form $S_{t,t'}\,=\,S(t-t')$.
Using the definition $S\,=\,W_+^{-1}W_-$, where
$$W_\pm\,=\,s-\lim_{\tau\to\pm\infty}U(-\tau)U_0(\tau)\,,$$
we find
$$(S\psi)_t\,=\,s-\lim_{\tau_1,\tau_2\to\infty}
(U_0(-\tau_1) U(\tau_1)U(\tau_2)U_0(-\tau_2)\psi)_t$$
But,
$$(U_0(-\tau_1)U(\tau_1 + \tau_2)
U_0(-\tau_2)\psi)_t\,=\,(U(\tau_1
+\tau_2)U_0(-\tau_2)\psi)_{t+\tau_1}\,=$$ $$=\,\int
W_{t+\tau_1,t'}(\tau_1+\tau_2)(U_0(-\tau_2)\psi)_{t'}dt'
\,=$$ $$=\,\int
W_{t+\tau_1,t'}(\tau_1+\tau_2)\psi_{t'+\tau_2}dt'\,=\,\int
W_{t+\tau_1,t'-\tau_2}(\tau_1+\tau_2)\psi_{t'}dt',$$ and
therefore the matrix elements of $S$ are
$$S_{t,t'}\,=\,s-\lim_ {\tau_1,\tau_2\to\infty}
W_{t+\tau_1,t'-\tau_2}(\tau_1+\tau_2)\,=$$ $$=\,s-\lim_
{\tau_1',\tau_2'\to\infty}
W_{t-t'+\tau_1',-\tau_2'}(\tau_1'+\tau_2')\,
=\,S(t-t')\eqno(2.11)$$
(where $\tau_1'=\tau_1+t'\,\,\tau_2'=\tau_2-t'$  ) . This
is a very important property of the $S$-matrix, according to
which, when one goes to the spectral representation
 ${\hat\psi}_\sigma\,=\,\int e^{-i\sigma t}\psi_tdt$, the
$S$-matrix takes the simple form
$$\hat S_{\sigma,\sigma'}\,=\,{1\over{2\pi}}\int\,
e^{-i\sigma t}
S_{t,t'}e^{i\sigma't'}dtdt'\,=\,\delta(\sigma-\sigma')\hat
S(\sigma)$$ where  $$ \hat S(\sigma)\,=\,\int e^{-i\sigma
t}S(t)dt , \eqno(2.12)$$ i.e., in this basis the $S$-matrix is
diagonal, and the $S$-operator is multiplication on the
subspaces, labeled by $\sigma$, of $\{{\cal
H}_{\sigma}\}$, the set of (isomorphic) Hilbert spaces
which are the Fourier dual to the set $\{{\cal H}_t\}$.
This result can be obtained also by looking at the
definition of the $S$-matrix, $$S\,=\,s-\lim_
{\tau_1,\tau_2\to\infty}U_0(-\tau_1) U(\tau_1 +
\tau_2)U_0(-\tau_2)$$ from which it follows that
$$SU_0(\tau)\,=\,U_0(\tau)S\,.$$ Since $U_0(\tau)$ is the
translation operator one obtains the result
$[\,S\,,\,i\partial_t\,]\,=\,0$ (which correspond to the
usual result of scattering theory
$[\,S\,,\,H_0\,]\,=\,0$). It follows from this commutation
relation that $S_{t,t'}\,=\,S(t-t')$.

\par We show now that under the  general
evolution (2.10), the semigroup is contractive.
 Let us calculate the
generator of the semigroup $B$ of ${\cal Z}(\tau) = P_+ U(\tau)P_- $.
We use the free
translation representation in which both ${\cal D}_\pm$ have
definite support properties. In this representation,  $${\cal
Z}(\tau)\,=\,P_+U(\tau)P_-\,=\,E(\rho)U(\tau)(I-E(0))
\,,\eqno(2.13)$$
where $E(t)$ is the spectral resolution corresponding to
$T_0$, the free-time-operator (the conjugate of $K_0$
which is, in the free translation representation, $-i\partial_t$ ).
Then, the generator (in the subspace $\cal K$) of ${\cal
Z}(\tau)$ is  $$B\,=\,i\, \lim_{\tau\to0}{{\cal
Z}(\tau)-I_{\cal K}\over\tau}\,=\, i\, \lim_
{\tau\to0}{E(\rho)(I-iK\tau)(I-E(0))-I_{\cal
K}\over\tau}\,=$$
$$=\,E(\rho)K(I-E(0))\,=\,P_+KP_-\,.\eqno(2.14)$$
 According to the requirements on
${\cal D}_\pm$ the matrix elements of $\kappa$,
 the self-adjoint kernel over $t,t'$ contained in $K$ distinct
 from the $t$-derivative [10],between
states from ${\cal D}_-$ to ${\cal D}_+$, or ${\cal
D}_{\pm}$ to ${\cal K}$ vanish, and therefore
$$B\,=\,P_+K_0P_-\,+\,\kappa_{\cal K}\,.\eqno(2.15)$$

An operator $B$ is called dissipative [17][18] if
$$-i\bigl( (\phi,B\phi)-(B\phi,\phi)\bigr)\leq 0\,,
\eqno(2.16)$$
for all $\phi$ in the domain of $B$. Since $\kappa_{\cal
K}$ is self-adjoint only the first term determines whether
the operator is dissipative, i.e., this property does not
depend on the perturbation. As shown by Horwitz and
Piron[14], the operator $P_+K_0P_-$ is, in fact,
dissipative. It is known [18] that ${\cal Z}(\tau)$ is
a contractive semigroup if and only if its generator is
dissipative. It therefore follows, independently of
(self-adjoint) interaction, that the semigroup ${\cal
Z}(\tau)$ is contractive. We see from this [14] the {\it essential
mechanism of Lax-Phillips theory}. The non-self-adjointness of
$P_+K_0P_-$ corresponds to the restriction of
$-i\partial_t$ to a finite interval, so that, in fact the
operator has imaginary eigenvalues. In the presence of
interaction (non-trivial $\kappa$), these eigenvalues
emerge as the actual eigenvalues of $B$, corresponding to
the singularities of $S(\sigma)$.

\par We remark that the direct integral space
provides a framework as a functional space for quantum mechanics in which
the Nagy-Foias construction can be realized, i.e., for which
unitary evolution can be restricted to a contractive semigroup.
 We shall now introduce
 an extension of the conceptual framework which considers
the set $\{\psi_t \}$, corresponding
 to the Lax-Phillips vector $\psi$, as
an ensemble of the same type, for
 example, as $\{\psi (x)\} \in {\cal H}$,
where $x$ is a point of the spectrum
 of the position observable, in the usual form
of the quantum theory.  In concluding
 this section, we investigate some
consequences of this interpretation.
\par In particular, we discuss some
 properties of the time operator and the realization
of the superselection rule in time. In the next section, we discuss the
possibility of decoherence in ${\cal H}$ induced by the unitary evolution in
${\bar {\cal H}}$.
\par
There are three distinct types of time operator. One,
which we call the incoming time operator $T^{in}$,
provides a spectral family in terms of which the incoming
representation can be constructed, and in which functions
in  ${\cal D}_-$ have definite support and functions in
${\bar{\cal H}}$ evolve by translation. In this
representation, the norm of the evolving states in
$(-\infty,0)$ must decrease. After sufficient
laboratory time $\tau$ passes, the states evolve to ${\cal
D}_+$, and in the outgoing representation, provided by the
spectral family of the outgoing time operator $T^{out}$,
they have definite support in $(\rho,\infty)$.
The mapping of functions in the incoming representation to
the outgoing representation is provided by the
Lax-Phillips $S$-matrix, and the time operators are
related by  $$T^{out}\,=\,ST^{in}S^{\dag}\,.\eqno(2.17)$$
The third type of time operator
corresponds to the ``free'' representation and
is related to $T^{in}$,$T^{out}$ by the Lax-Phillips wave
operators. The spectral family for this operator provides
the ``standard'' representation
 (analogous to Dirac's choice of ``standard''
spectral families), which we have used above.

There is an interval, in general, when the system is in
interaction, and its state is neither in ${\cal D}_-$ nor
${\cal D}_+$. The expectation value of the operator
$T^{in}$ in the state $\psi^{\tau}$ projected into ${\cal
K}\oplus{\cal D}_+$ (corresponding to the projection
$P_-$) can be interpreted as the interaction interval. If
the system in interaction is considered as an unstable
particle (a resonance), this interval is its {\it age}
after creation at $t=0$.
 The expectation value of $T^{in}$ then moves out
of $(-\infty,0)$. The expectation value of $T^{in}$ in the
state $P_-\psi^{\tau}$ is  $$<T^{in}>_{\tau}\,=\,\int t |\
{}_{in}\langle t|P_-\psi^{\tau})|^2dt\,;\eqno(2.18)$$  here,
$|\ {} _{in}\langle t|P_-\psi^{\tau})|^2$ is the probability
density for the age $t$ at time $\tau$, an intrinsic
dynamical property of the system. The positive value that
the expectation value develops corresponds to the average
age. One can similarly compute the expected time after decay,
the expected lifetime, and the expected value of any other
observable of interest as a property of the unstable system.
\par  We then understand the
subspace ${\cal K}$ as corresponding to  the unstable
system.
\par  The structure of the theory is somewhat similar to
the Wigner-Weisskopf idea, in that a subspace is
associated with the decaying system. The decay of the
system is associated with the probability flow out of
the subspace. As in the original Wigner-Weisskopf formulation, the
process of decay may be represented as a continuous evolution
from the original unstable state to the final state
through a changing linear superposition. In this framework, let
us choose a vector $\psi$ in the subspace $\cal K$ to
represent the state of an unstable system. Then, under the
full evolution,
$$\eqalign{(\psi, U(\tau)\psi)&=
 (\psi, P_{\cal K} U(\tau) P_{\cal K} \psi)\cr
                     &= (\psi, {\cal Z}(\tau) \psi), \cr}
\eqno(2.19) $$
so that {\it the reduced evolution is an exact semigroup}.
\par  Moreover, in the
Lax-Phillips theory the expectation value of an observable
which is decomposable in the free or outgoing
representations, where ${\cal D}_+$ has definite support
properties, necessarily reduces to the sum of the
expectation values in the subspaces ${\cal K}\oplus{\cal
D}_-$ and in the subspace ${\cal D}_+$ (the decay
products), i.e.,
$$ \langle A \rangle = \int \, dt (\psi_t,\, A\psi_t) =
\sum_{M=D_{\pm},{\cal K}} \int \, dt \, (\psi_t^M,\, A \psi_t^M ).
\eqno(2.20)$$
\par Note that there are no cross terms.
 {\it There is, therefore, an exact
superselection rule for measurements of the system by
means of such decomposable operators.}

\bigskip
\noindent {\bf 3. APPLICATIONS}
\smallskip
\noindent {\it a. Measurement according to Namiki and Machida}
\smallskip
\par Recently, Machida and Namiki [19] have proposed a
measurement theory based on a direct integral space of
continuously many Hilbert spaces and a continuous
superselection rule. As pointed out by Tasaki {\it et
al} [20], although they had some success, their theory has
a conceptual difficulty. Indeed, in their theory, while
the apparatus is described by many Hilbert spaces, the
system corresponds to a single Hilbert space as in the
conventional theory. Thus, one needs to specify the
boundary between the system and the apparatus. As
discussed by von Neumann, this is impossible.

\par  Most measurement processes are
 concerned with measurements of observables
which are time-independent in
 the Schr\"odinger picture.  Therefore, if two different
Lax-Phillips states give the
 same expectation value for all time-independent
observables, these two states
 are essentially indistinguishable.  In this sense, we
define the following:
\smallskip
\noindent\item{1.} A Lax-Phillips
 vector $\psi \in {\bar{\cal H}}$ is called `` effectively pure''
if there exists a pure
 state $$\rho_0 = \phi_0 \phi_0^*, \qquad
  \phi_0 \in {\cal H},$$  such that
$$ \langle {\hat A}
 \rangle_\psi = {\rm Tr} \rho_0 A = ( \phi_0, A \phi_0), \eqno(3.1) $$
where ${\hat A}$ is the ``lift'' of $A$ on $\cal H$ to ${\bar{\cal H}}$,
for every element of
 the algebra of bounded linear operators associated with the
spectral families of
 the time-independent observables\footnote{*}{ We wish to emphasize
that what is meant is
 {\it explicit} time-dependence in the Schr\"odinger picture; we
do not refer here to the
 dynamical time-dependence that may arise in the Heisenberg
picture if $A$ is not a
 constant of the motion.} on the original space ${\cal H}$.

\noindent
\item {2.} A Lax-Phillips vector
 is called ``effectively mixed'' if no such (pure) $\rho_0$ exists.

It can be shown[10][20]  that $\psi = \{\psi_t\}
 \in {\cal H}$ is effectively pure if an only if
it has the form
$$ \psi_t = f(t) \phi_0 . \eqno(3.2)$$
\par We now discuss the possibility of decoherence, or the
evolution from effectively pure to effectively mixed states. First, we
consider the Schr\"odinger evolution for a time-dependent
Hamiltonian.
 The solution of the time-dependent Schr\"odinger equation can
always be written formally as $\psi_t=U(t,t')\psi_{t'}$,
where $U(t,t')$ satisfies the chain property
$U(t,t')U(t',t'')=U(t,t'')$, and can be expressed in
terms of the integral of a time-ordered product. We
define $W_t(\tau)=U(t+\tau,t)$, and lift the evolution to
$\bar{\cal H}$ as follows
$$\psi^{\tau}_{t+\tau}\,=\,W_t(\tau)\psi_t\,,\eqno(3.2)$$
where $W_t(\tau)$ is given by ($T$ implies the
time-ordered product)
$$W_t(\tau)\,=\,T\left(e^{-i\int_t^{t+\tau}H(t')dt'}
\right)\,.\eqno(3.3)$$
 For this kind of time-evolution
we obtain
$$ \langle {\hat A} \rangle_\psi
 = \int dt\ (W_t(\tau) \psi_t, A W_t(\tau)\psi_t)_
{\cal H} , \eqno(3.4) $$
where we have taken the
 normalization as unity.  For the effectively pure states
 we have
$$ \langle {\hat A} \rangle_\psi = \int dt\ \vert f(t)\vert^2 (W_t(\tau)
\phi_0, A W_t(\tau) \phi_0)_{\cal H} . \eqno(3.5)$$
It follows from our
 previous argument
  that the effective state corresponding to (3.5)
is mixed-like if $W_t(\tau) \phi_0 \neq W_{t'}
(\tau) \phi_0$ (i.e., the state $\rho_\psi$
 induced from $\psi^\tau_{t+\tau} =
W_t(\tau) \psi_t = f(t) W_t(\tau) \phi_0 $ is not pure in ${\cal H}$).
This result is true for the generalized evolution  $W_{tt'}$ of
$(2.10)$ as well.
\bigskip
\noindent {\it b. Intrinsic decoherence in classical and quantum
Liouville evolution.}
\smallskip
\par It has long been emphasized
 by Prigogine and his co-workers [21] that the natural
description for the evolution of a
 system with many degrees of freedom is that of the
evolution of the density matrix
 $\rho$, through the Liouville equation,
   $$i{d\rho \over
dt}\,=\,[\,H\,,\,\rho\,]\,. \eqno (3.6)$$
 The density matrix $\rho$ ($ \rho
\geq 0,Tr\rho=1$) has the property that $Tr\rho^2
\leq 1$, where the equality is attained
 only for a pure state. In general, one considers
the space of Hilbert-Schmidt operators
 $A$ for which  $$ Tr\, A^*A\,<\,\infty\,; \eqno
(3.7)$$ the positive (normalized)
 elements of such a space correspond to the physical
states, the density matrices.
 On this space, the commutator with the Hamiltonian $H$
defines a linear operator $\cal L$ , called the Liouvillian, for which
$$i{d\rho \over d\tau}\,=\,{\cal L}\rho\, ,
  \eqno(3.8)$$ where one assumes that $\cal
L$ is self-adjoint in the Liouville space.
\par The Hamiltonian evolution of states in classical mechanics is known
by the Liouville theorem to be non-mixing, i.e., to preserve the
entropy of the system [22].  The same property holds for the
quantum evolution as well, and follows from the unitarity of the
evolution operator.  This has been an obstacle to the consistent
description of irreversible processes from first principles [23].
The usual use of techniques of coarse graining or truncation
to achieve a realization of the second law does not follow from
basic dynamical laws, and is fundamentally not consistent with
the underlying
Hamiltonian structure [24].  We shall now show that the existence
of a time operator in the Liouville space provides a natural
and consistent mechanism for the decoherence of physical states,
i.e., that pure states become mixed during the evolution, both for
quantum and classical systems.

In particular, for a Hamiltonian
 of the form of the sum of an unperturbed operator $H_0$
and a perturbation $V$, i.e.,
$H\,=\,H_0\,+\,V$ , the corresponding Liouvillian is
$$ {\cal L}\,=\,{\cal L}_0\,+\,{\cal L}_I\,.
  \eqno(3.9)$$ Now suppose we consider the
``time operator'' $T_0$, conjugate to  ${\cal L}_0$
(with spectrum $(-\infty, \infty)$; it satisfies
$$[\,T_0\,,\,{\cal L}_0\,]\,=\,i\,.$$ Then,
 in the spectral representation of $T_0$,  $$ {}_0\langle
t|[T_0,{\cal L}_0]|t'\rangle_0\,=\,i\delta(t-t')\,,$$ or
$$(t-t')\ _0\langle t|{\cal L}_0|t'\rangle_0\,=\,i
\delta(t-t')\,. \eqno(3.10)$$  It follows that
$$ {}_0\langle t|{\cal L}_0|t'\rangle_0\,=\,-i\partial_t
\delta(t-t')\,. \eqno (3.11)$$  Hence,
$$ {}_0\langle t|{\cal L}|t'\rangle_0\,=\,-i\partial_t\delta(t-t')\,+
 \, _0
\langle t|{\cal L}_I|t'\rangle_0\,,
 \eqno(3.12)$$  where the last term
  is, in general,
not diagonal.
\par The method that we have described above applies as well
to the formulation of classical mechanics on a Hilbert
space defined on the manifold of phase space which was
introduced by Koopman [25]  and used extensively in
statistical mechanics [24]. Misra [26] has shown that
dynamical systems which admit a Lyapunov operator
necessarily have absolutely continuous spectrum in $(-\infty,\infty)$;
therefore one can construct a time operator on the
classical Liouville space for such systems.
 The expectation value of a
$t$-independent operator defines a reduced density
function in the form
$$ \int dt\ \rho_t(\beta), $$
where $\beta$ is the set of variables remaining after extracting
$t$ as a function on the manifold of the measure (phase) space.
Since a pure state is
defined by a density function concentrated at a point
of the phase space, a state which is effectively pure must
have the form $\delta(\beta-\beta_0)$. The
equivalence class associated with this reduced density
contains mixed states as well, such as
$\rho(t,\beta)=\delta(\beta-\beta_0) f(t)$
corresponding to a non-localized function on the phase
space. The structure of the
theory, and the conclusions we have reached, are
therefore identical to those of the quantum case.
\bigskip
{\it c. Relativistic quantum mechanics.}
\smallskip
\par The form of relativistic quantum mechanics introduced
by Stueckelberg [27], extended to the many-body case by
Horwitz and Piron [28], covariantly describes the evolution of
a system according to the Stueckelberg-Schr\"odinger equation
$$ i{\partial \psi_\tau \over \partial \tau} = {p_\mu p^\mu \over
2M} \psi_\tau \equiv K_0 \psi_\tau, \eqno(3.13)$$
where $M$ is an intrinsic property of the particle (``on-shell''
mass).
The classical form of this theory has for its Hamilton equations
$$ {dx^\mu \over d\tau} = {\partial K_0 \over \partial p_\mu}
= {p^\mu \over M}, \eqno(3.14)$$
and, therefore, eliminating $d\tau$, one obtains the standard
relativistic relation
$$ {d{\bf x} \over dt} = {{\bf p}\over E} . \eqno(3.15)$$
Since the d'Alembertian, corresponding to the operator $K_0$,
has spectrum $(-\infty, \infty)$, there exists an operator
$\xi$ which satisfies
$$ [K_0,\, \xi] = i. \eqno(3.16)$$
Note that the operator $t$ of the relativistic theory will not
serve this purpose, since its commutator with $K_0$ is
$i E/M $, which only approaches $i$ in the non-relativistic limit.
\par If $\xi$ is a function of ${\bf x},t$, we may construct
the transformation function $\langle \xi',\beta \vert x \rangle$
using the defining commutation relation, i.e.,
$$ \langle \xi',\beta \vert K_0 \xi - \xi K_0 \vert x \rangle
= i \langle \xi', \beta \vert x \rangle,$$
or
$$ i {\partial \over \partial \xi' } \bigl( \xi' \langle \xi',\beta
\vert x \rangle \bigr) - \xi' \bigl(-{\partial^\mu \partial_\mu
\over 2M} \bigr) \langle \xi', \beta \vert x \rangle = i \langle
\xi', \beta \vert x \rangle,$$
so that we obtain the defining equation [29]
$$  i {\partial \over \partial \xi'} \langle \xi', \beta
\vert x \rangle = -{\partial^\mu \partial_\mu \over 2M}
\langle \xi', \beta \vert x \rangle. \eqno(3.17)$$
\par We thus see that the relativistic quantum theory provides a
natural framework for the Lax-Phillips formulation of the description
of an unstable system.  It is interesting that the continuous
spectrum of $K_0$ is essential to the construction; this implies
that we must have both positive and negative mass-squared states
in the spectrum, i.e., that the so-called tachyons,
at least in the form of intermediate states, play a fundamental
role in the relativistic description of unstable systems.
\par One might ask how such a mechanism could survive in the
non-relativistic limit.  It is interesting to study this
limit; even though the Galilean world is an idealization which
is not realized physically, the velocity of light $c$ is very large.
To study this limit, we consider the condition
$$ E - Mc^2 = \varepsilon < \infty , \eqno(3.18)$$
for $c \rightarrow \infty$,
used, for example, in ref. [30].  We then define the variable $m$
such that
$$ E =c\sqrt{p^2 + m^2 c^2}, \eqno(3.19)$$
so that
$$ E-M c^2 = (m-M)c^2 + mc^2 \bigl\{ \sqrt{1+ {p^2 \over m^2c^2}}
-1 \bigr\}. \eqno(3.20)$$
Defining
$$\eta = (m-M)c^2  \neq 0,$$
we see that
$$ E = Mc^2 + \eta + {p^2 \over 2M} + f(\eta, p^2), \eqno(3.21)$$
an integral kernal on the wave functions, where the integral
operator is ${\rm O}(1/c^2)$ .
\par The general structure of the relativistic
Lax-Phillips theory therefore
remains in the Galilean limit (for finite but large $c$). The
experimental signature of such an additional term in the Hamiltonian
would be, for example, an interference effect in time.  The
experiment would be of the same design as the test of such an
effect in the full covariant relativistic theory [31].
\bigskip

\noindent{\it Acknowledgements}
\smallskip

One of us (LPH) wishes to thank C. Piron for many discussions,
and he is grateful for his hospitality in Geneva
during several visits.
 We wish to thank our colleagues in Brussels,
 I. Antoniou, B. Misra, I. Prigogine, and S. Tasaki,
 for  helpful discussions and comments
  which contributed significantly to many
of the results that we have presented.
\bigskip
\frenchspacing
\noindent {\bf References:}
\item{1.} G. Gamow, Zeits. f. Phys. {\bf 51},
            204 (1928).
\item{2.} T.D. Lee, R. Oehme and C.N. Yang, Phys.
       Rev. {\bf 106}, 340 (1957); T.T. Wu and
        C.N. Yang, Phys. Rev. Lett. {\bf 13}, 380
      (1964).
\item{3.} V.F. Weisskopf and E.P.Wigner, Zeits. f.
     Phys. {\bf 63}, 54 (1930); {\bf 65},
                                     18 (1930).
\item{4.} L.P. Horwitz and J.P. Marchand, Helv.
           Phys. Acta, {\bf 42}, 1039 (1969);
 L.P. Horwitz and J.P. Marchand, Rocky Mtn.
             Jour. Math. {\bf 1}, 225 (1973).
\item{5.} K.O. Friedrichs, Comm. Pure Appl. Math.
 {\bf 1}, 361 (1948);T.D. Lee, Phys. Rev. {\bf 95}, 1329 (1956).

\item{6.} N. Bleistein, H. Neumann, R, Handelsman
   and L.P. Horwitz,
     Nuovo Cim. {\bf 41A},   389 (1977).
\item{7.} C. Piron, {\it Foundations of Quantum Physics},
Benjamin/Cummings, Reading, Mass. (1976).
\item{8.} L.P. Horwitz, J.P. Marchand and J. LaVita, Jour. Math.
Phys. {\bf 12}, 2537 (1971); D. Williams, Comm. Math. Phys.
{\bf 21}, 314 (1971).
\item{9.} L.P. Horwitz and I.M. Sigal, Helv. Phys.
                   Acta {\bf 51}, 685 (1980);
W. Baumgartel, Math. Nachr. {\bf 75}, 133 (1978).  See also,
G. Parravicini, V. Gorini
and E.C.G. Sudarshan, Jour. Math. Phys. {\bf 21}, 2208 (1980);
A. Bohm, {\it The Rigged Hilbert Space and Quantum Mechanics}, Springer
Lecture Notes on Physics {\bf 78}, Berlin (1978);
 A. Bohm, {\it Quantum Mechanics:
Foundations and Applications},
 Springer, Berlin (1986); A. Bohm, M. Gadella
and G.B. Mainland, Am. Jour. Phys. {\bf 57}, 1103 (1989); T. Bailey and
W.C. Schieve, Nuovo Cimento {\bf 47A}, 231 (1978).
\item{10.} E. Eisenberg and L.P. Horwitz, {\it Advances in
Chemical Physics}, to be published (1996).
\item{11.} P.D. Lax and R.S. Phillips, {\it Scattering
         Theory}, Academic Press, New York (1967).
\item{12.} See for example: L.E. Reichl, {\it
        The Transition to Chaos in Conservative
         Classical Systems: Quantum Manifestation},
         Springer, New York, (1992), and references
     therein; I.M. Sigal, personal communication.
\item{13.} C. Flesia and C. Piron, Helv. Phys. Acta
                        {\bf 57}, 697 (1984).
\item{14.} L.P. Horwitz and
 C. Piron, Helv. Phys. Acta {\bf 66}, 694 (1993).
\item{15.} I.P. Cornfeld, S.V. Fomin and Ya.G Sinai,
{\it Ergodic Theory}, Springer, Berlin (1982).
\item{16.} C. Foias and B.Sz. Nagy, Acta Sci.
        Math. {\bf 23}, 106 (1962); F. Riesz and
        B.Sz. Nagy, {\it Functional Analysis},
         Appendix by B. Sz. Nagy, 2$^{nd}$
            edition. English Trans. Dover,
                               New York (1990).
\item{17.} M. Reed and B. Simon, {\it Methods of
       Modern Mathematical Physics}, vol. 3, p.
            236, Academic Press, New York (1979).
\item{18.} E.B. Davies, {\it Quantum Theory of Open Systems},
p. 103, Academic Press, London (1976).
\item{19.} Machida and Namiki, Prog. Theor. Phys.,
      {\bf 63}, 1457 (1980); {\bf 63}, 1833
       (1980); Namiki and Pascazio, Phys. Rev.
                        {\bf A44}, 39 (1991).
\item{20.} S. Tasaki, E. Eisenberg and L.P. Horwitz,
           Found. of Phys. {\bf 24}, 1179 (1994).
\item{21.} C. George, Physica {\bf 65}, 277 (1973); I.
          Prigogine, C. George, F. Henin and L.
           Rosenfeld, Chemica Scripta {\bf 4}, 5
         (1973); T. Petrosky, I. Prigogine and S.
           Tasaki, Physica {\bf A173}, 175 (1991);
          T. Petrosky and I. Prigogine, Physica
          {\bf A175}, 146 (1991), and references
                                     therein.
\item{22.} See, for example, J. Yvon,  {\it
       Correlations and Entropy in Classical
       Statistical Mechanics}, Pergamon Press,
                             Oxford (1969).
\item{23.} See, for example, I.E. Antoniou and I.
       Prigogine, Physica {\bf A192}, 443 (1993).
\item{24.} I. Prigogine, {\it Non-equilibrium
      Statistical Mechanics}, Wiley, New-York (1961).
\item{25.} B. Koopman, Proc. Nat. Aca. Sci. {\bf 17},
                                           315 (1931).
\item{26.} B. Misra, Proc. Nat. Aca. Sci. {\bf 75},
         1627 (1978); B. Misra, I. Prigogine and
          M. Courbage, Proc. Nat. Aca. Sci. {\bf
                           76}, 4768 (1979).
\item{27.} E.C.G. Stueckelberg, Helv. Phys. Acta {\bf 14}, 372,
588 (1941); {\bf 15}, 23 (1942).
\item{28.} L.P. Horwitz and C. Piron, Helv. Phys. Acta {\bf 46},
316 (1973).
\item{29.} Y. Strauss and L.P. Horwitz, in preparation.
\item{30.} L.P. Horwitz, W.C. Schieve, and C. Piron, Ann. Phys.
{\bf 137}, 306 (1981).
\item{31.} L.P. Horwitz and Y. Rabin,  Lett. Nuovo Cimento {\bf 17},
501 (1976).

\vfill
\end
\bye